\documentclass[aps,prc,showpacs,floatfix,footinbib,twocolumn]{revtex4-1}
\usepackage{amsmath}
\usepackage[dvips]{graphics,graphicx}
\usepackage{graphicx}
\usepackage{dcolumn}
\usepackage{bm}

\begin{document}

\title{Thermodynamics of a gas of hadrons with attractive and repulsive
interaction within S-matrix formalism}

\author{Ashutosh Dash}
\email{ashutosh.dash@niser.ac.in}

\author{Subhasis Samanta}
\email{subhasis.samant@gmail.com}

\author{Bedangadas Mohanty}
\email{bedanga@niser.ac.in}
\altaffiliation[]{On sabbatical leave to Experimental
Physics Department, CERN, CH-1211, Geneva 23, Switzerland}

\affiliation{School of Physical Sciences, National Institute of Science 
Education and Research, HBNI, Jatni - 752050, India}

\begin{abstract}
We report the effect of including repulsive interactions on various thermodynamic observables 
calculated using a S-matrix based Hadron Resonance Gas (HRG) model to already available 
corresponding results with only attractive interactions \cite{Dash:2018can}. 
The attractive part of the interaction is calculated by parameterizing the two body phase shifts
using K-matrix formalism while the repulsive part is included by fitting to the experimental phase 
shifts which carry the information about the nature of the interaction. We find that the bulk
thermodynamic variables for a gas of hadrons such as energy density, pressure, entropy 
density, speed of sound and specific heat are suppressed by the inclusion of repulsive
interactions and are more pronounced for
second and higher order correlations and fluctuations, particularly for the observables $\chi^2_Q$,
$\chi^2_B-\chi^4_B$ and $C_{BS}$ in the present model. We find a
good agreement between lattice QCD simulations and the present model for $C_{BS}$. We have also computed
two leading order Fourier
coefficients of the imaginary part of the first order baryonic susceptibility at imaginary baryon chemical potential
within this model and compared them with the corresponding results from lattice.
Additionally, assuming that the 
value of interacting pressure versus temperature for a gas of hadrons calculated in S-matrix 
formalism is same as that from a van der Waals HRG (VDWHRG) model, we have quantified the 
attractive and repulsive interactions in our model in terms of  attractive and repulsive 
parameters used in the VDWHRG model. The values of parameters thus obtained are $a=1.54\pm 0.064$~GeV $\text{fm}^{3}$ and $r=0.81\pm 0.014$~fm.
\end{abstract}

\maketitle

\section{\label{sec:Intro}Introduction}

One of the primary goals of relativistic heavy ion collision is the study of QCD 
(Quantum Chromo Dynamics)
phase diagram \cite{Andronic:2008gu}.
There are at least two phases in the phase diagram, one where the degree of freedom are
quarks and gluons called the Quark Gluon Plasma (QGP) phase and other where the degrees 
of freedom are hadronic. An approach to study the properties of hadronic phase formed by
hadronization of the QGP is through a statistical model of a gas of hadrons called the 
hadron resonance gas model (HRG) \cite{Becattini:2009sc}.
The hadron resonance gas (HRG) ~\cite{BraunMunzinger:1994xr, Cleymans:1996cd, Yen:1998pa,
BraunMunzinger:1999qy, Cleymans:1999st,Becattini:2000jw, BraunMunzinger:2001ip, Becattini:2005xt,
Andronic:2005yp, Andronic:2008gu, Das:2016muc,Andronic:2017pug} models have successfully
described the hadron multiplicities produced in relativistic nuclear collisions
over a wide range of center of mass energies. The main result of such an investigation was the
observation of rise in the extracted
chemical freeze-out temperature values from lower energies to almost a constant 
value of temperature $T \simeq 155-165$~MeV at higher energies, supplemented with the decrease 
of the baryon chemical potential ($\mu_B$) with increasing energy \cite{BraunMunzinger:2007zz}. 
The saturation of temperature supports the Hagedorn's limiting temperature hypothesis 
\cite{Hagedorn:1965st}, suggesting the possibility of a phase boundary. Similarly, 
theoretical investigation of QCD on lattice (LQCD) at vanishing $\mu_B$ indeed predicts
a sharp increase of thermodynamical quantities near deconfinement temperature 
$T_c$ ~\cite{Aoki:2006we,Borsanyi:2011sw,Gupta:2011wh, Bazavov:2012jq,Bellwied:2013cta, Borsanyi:2013bia, Bazavov:2014pvz, Bellwied:2015lba}. 
The HRG model is also successful in describing LQCD data related to the 
bulk properties of hadronic matter in thermal and chemical equilibrium below $T_c$ \cite{Borsanyi:2011sw, Bazavov:2012jq, Bazavov:2014pvz,
Bellwied:2013cta, Bellwied:2015lba,Karsch:2003zq}.

\par
The phenomenal success of the ideal HRG (IDHRG) model in predicting the hadronic yields 
can be attributed to a theorem by Dashen and Ma \cite{Dashen:1969ep} which states that 
the  partition function of an interacting hadronic gas, can be decomposed into a free 
and an interacting part. Considering that only resonances contribute to the interacting 
part, it can be shown that in a narrow resonance width approximation, the net effect of the 
interacting part is equivalent to considering all such hadronic resonances as free particles. 
However, relaxing the above assumptions by including resonances of 
finite widths (both overlapping and non-overlapping), it has been seen that the variation of thermodynamic variables with temperature
changes substantially \cite{Gorenstein:1987zm,Venugopalan:1992hy,Weinhold:1997ig,Wiranata:2013oaa,Huovinen:2016xxq}. 
Further, it can be argued that such interaction contribute only to the attractive part of partition function and the inclusion of a
repulsive part could partially negate the effect of the attractive 
part. For example, in Refs.~\cite{Rischke:1991ke, Cleymans:1992jz, Yen:1997rv, 
Tiwari:2011km,Begun:2012rf, Andronic:2012ut, Fu:2013gga, Bhattacharyya:2013oya,Albright:2014gva,
Vovchenko:2014pka,Albright:2015uua,Kadam:2015xsa,Kapusta:2016kpq,Satarov:2016peb,Vovchenko:2016ebv,Adak:2016jtk,
Alba:2016fku,Alba:2016hwx,Vovchenko:2018cnf} the authors have used an excluded volume approach which only 
had the repulsive part whereas Refs.~\cite{Vovchenko:2015xja, Vovchenko:2015vxa, Vovchenko:2015pya, 
Redlich:2016dpb,Vovchenko:2016rkn,Vovchenko:2017cbu,Vovchenko:2017zpj,Samanta:2017yhh} 
considered an  van der Waals' (VDW) type of interaction, 
which has both the attractive and repulsive part and a comparison of the calculated 
thermodynamic pressure from both the approaches shows the feature as discussed above.

\par
In our previous work Ref.~\cite{Dash:2018can}, we had developed a HRG model with attractive interactions
between hadrons using the K-matrix formalism. In the present work, we extend the K-matrix formalism to include 
repulsive interactions between the hadrons using the S-matrix formalism. 
In Ref.~\cite{Dash:2018can}, we used K-matrix formalism to calculate the phase shifts of the resonance spectral function 
in contrast to the popular Breit-Wigner parametrization. It has been argued previously that the K-matrix
formalism preserves the unitarity of the scattering matrix (S-matrix) and neatly handles multiple
resonances \cite{Chung:1995dx,Wiranata:2013oaa,Dash:2018can}. However, the formalism fails to handle any repulsive channel in the scattering matrix. 
Therefore in this work, we include the repulsive part by fitting to experimental phase shifts that 
encodes the information about the nature of interaction. We use the phase shifts data from Scattering 
Analysis Interactive Database (SAID) partial wave analysis for nucleon-nucleon ($N N$), pion-nucleon
($\pi N$) and kaon-nucleon ($K N$) interaction in their respective isospin channels \cite{Workman:2016ysf,Workman:2012hx,Hyslop:1992cs}. Additionally, 
we have also included the repulsive isotensor channel in the pion-pion ($\pi\pi$) scattering, as has 
been pointed in many earlier works \cite{Venugopalan:1992hy,Broniowski:2015oha}.

After constructing the interacting hadron resonance gas model with both attractive and
repulsive interactions using phase shift information for various hadronic interactions 
we calculate the various thermodynamic observables like pressure, energy density, entropy
density, interaction measure,  specific heat, speed of sound and susceptibilities. 
The temperature dependence of these observables are then compared with corresponding 
results from Lattice QCD, IDHRG and HRG models with attractive interactions using K-matrix formalism.

The paper is organized in the following manner. In the next section we discuss the 
formalism used to introduce repulsive interactions to our HRG model  developed earlier
using K-matrix approach with attractive interactions \cite{Dash:2018can}.
In Sec. \ref{sec:Result} we discuss 
the results from the new interacting HRG model with both attractive and repulsive 
interactions among the hadrons. The temperature dependence of our results are compared
to those from LQCD and IDHRG (with different hadron spectrum). Finally in 
Sec. \ref{sec:Summary} we summarize our findings. 

\section{\label{sec:Formalism}Formalism}
The equation of state for an interacting gas of hadrons of a single species can be computed by using the method 
of virial expansion. Specifically, the pressure of such a gas can be written 
as \cite{Landau:1980mil},
\begin{equation}\label{Eq:PressVirialExpansion}
 P(\beta,\mu)=\frac{1}{\beta}\sum_{i=1}^{\infty}\frac{J_i}{i!}\xi^i,
\end{equation}
where $\xi={\left(m/2\beta\pi\right)}^{3/2}e^{\beta \mu}$ and the inverse temperature, chemical potential, mass are denoted by $\beta,\mu, m$ respectively. 
The term $J_i$ takes into account the interaction between groups of $i$ hadrons and which are given as,
\begin{equation}
 J_1=1,\hspace{1cm} J_2=\int dV_2 \left(e^{-\beta U_{12}}-1\right),
\end{equation}
etc., where $U_{12}$ is the interaction energy. Differentiating Eq.~(\ref {Eq:PressVirialExpansion}) with respect
to $\mu$, we obtain the expression for number density i.e.,
\begin{equation}\label{Eq:NumberVirialExpansion}
 n(\beta,\mu)=\left(\frac{\partial P}{\partial \mu}\right)_{\beta,V} =\sum_{i=1}^{\infty}\frac{J_i}{(i-1)!}{\xi}^i.
\end{equation}
Eliminating $\xi$ to the first order from Eq.~(\ref{Eq:PressVirialExpansion}) and (\ref{Eq:NumberVirialExpansion}) gives us the ideal equation of state $P=nT$, where $T$ is the temperature. 
For a relativistic non-interacting quantum gas the expression for the pressure is given in \cite{Venugopalan:1992hy}. The classical virial equation of
state truncated at the second order is given as, $P=nT(1+nB(T))$, where $B(T)=-J_2/2$ is called the second virial coefficient.
In this work while calculating the virial coefficients we will be using the S-matrix approach to 
statistical mechanics, which has also been used previously in Refs.~\cite{Venugopalan:1992hy,Lo:2017lym,Lo:2017ldt,Huovinen:2017ogf} to study the thermodynamics of interacting hadrons.\par
In the S-matrix formalism, the second virial coefficient is related to the scattering amplitude or alternatively to the scattering phase shifts ${\delta}_l^I$ for a given spin $l$ and isospin $I$ 
channel. The correction to the ideal pressure for binary interactions between particles of species $i$ with particles of species $j$ is given as
\begin{align}
\begin{split}\label{Eq:IntPressure}
\operatorname{P_{int}^{ij}}&=\frac{T J_2}{2}z_i z_j\\
&=\frac{z_i z_j}{2{\pi}^3{\beta}^2}\int_{M_{ij}}^{\infty} d\varepsilon {\varepsilon}^2K_2(\beta\varepsilon)\sum_{I,l}{}^{'}g_{I,l}
\frac{\partial {\delta}_l^I(\varepsilon)}{\partial \varepsilon},
\end{split}
 \end{align}
where the terms $z_i$, $g_{I,l}$ and $\varepsilon$ stand for the fugacity, the spin-isospin
degeneracy factor and the total center of mass energy respectively. The function $K_2(x)$ 
stands for the modified Bessel function of second kind and the term $M_{ij}$ is the invariant mass 
of the interacting hadron pair $ij$ at threshold. Additionally, there is a sum over all possible spin-isospin 
channels and the prime over the summation sign denotes that for given $l$, the sum over $I$ is 
restricted to values consistent with statistics. A similar expression for interacting part of number density 
can also be derived such that $n_{\text{int}}^{ij}=\beta\operatorname{P_{\text{int}}^{ij}}$ for $i\neq j$ and 
$n_{\text{int}}^{i}=2\beta\operatorname{P_{\text{int}}^{i}}$ for $i=j$. \par

The total pressure and number density for an interacting system containing $N$ such hadronic species is then given as
\begin{eqnarray}
  \operatorname{P}=\sum_{i}^{N}\operatorname{P_{id}^{i}}+\sum_{i,j\geq i}^{N}\operatorname{P_{int}^{ij}}=\operatorname{P_{id}}+\operatorname{P_{int}},\\
  n=\sum_{i}^{N}{n_{\text{id}}^i}+\sum_{i,j\geq i}^{N}{n_{\text{int}}^{ij}}={n_{\text{id}}}+{n_{\text{int}}},
\end{eqnarray}
where $\operatorname{P}_{\text{id}}^i$, $n_{\text{id}}^i$ are the ideal contribution of the species $i$ to pressure and number density of the system respectively. 
In the present work, the contribution to the non-interacting (ideal) part comes from all the stable hadrons.
An equation of state truncated to the second
order can be derived as in previous paragraph such that the total pressure $P$ or the total number density $n$ is given as
\begin{eqnarray}\label{Eq:CanonPressure}
 \operatorname{P}&=&nT-\operatorname{P_{int}}{\left(\frac{n}{n_{\text{id}}}\right)}^2
 = nT+\operatorname{\overline{P}_{int}},
\end{eqnarray}
\begin{eqnarray}\label{Eq:Canonnumdensity}
n&=&\frac{P}{T}+\frac{n_{\text{int}}}{2}{\left(\frac{P}{P_{\text{id}}}\right)}^2
 =\frac{P}{T}+\overline{n}_{\text{int}},
\end{eqnarray}
 where
$\operatorname{\overline{P}_{int}}$ and $\overline{n}_{\text{int}}$ are the effective contribution of interaction 
to pressure and number density respectively.

\par
From Eq.~(\ref{Eq:IntPressure}), it can be seen that the second virial coefficient gives
positive (attractive) or negative (repulsive) contribution depending on whether the derivative
of phase shifts are positive or negative. The phase shifts are obtained from experiments or 
from theoretical calculations. In the present work, we determine the attractive phase shifts 
using the K-matrix formalism which takes the masses and partial widths of resonances from the 
PDG (Particle Data Group) \cite{Patrignani:2016xqp} as input.
Since the K-matrix formalism is not applicable for handling the repulsive phase shifts, these are obtained by fitting to experimental data. 
We would like to note here that since we do not have the information of masses and widths of resonances (mentioned in PDG) that decay into a pair of nucleons,
we extract phase shifts in such situation by fitting to experimental data.

\subsection{K-matrix Formalism}
A theoretical way of calculating phase shifts is to use the K-matrix formalism. 
The K-matrix formalism preserves the unitarity of S-matrix and neatly handles multiple
resonances \cite{Chung:1995dx}. In addition to that, widths of the resonances are handled 
naturally in the above formalism. In contrast, to the notion of ideal HRG is only valid for narrow
resonances and not for broad resonances, the K-matrix formalism can be applied quite generally.
Similarly, for overlapping resonances the K-matrix gives a more accurate description of the
phase shifts than the Breit-Wigner parametrization. In Ref.~\cite{Wiranata:2013oaa} the K-matrix
formalism was used to study an interacting gas of hadrons and it was
extended further in \cite{Dash:2018can}.  \par 
The resonances contributing to the process $ab\rightarrow R\rightarrow cd$, appear as a sum of poles in the K-matrix,
\begin{equation}\label{Eq:KmatrixkeyEqn}
 K_{ab\rightarrow cd}=\sum_{R}\frac{g_{R\rightarrow ab}(\sqrt{s}) g_{R\rightarrow cd}(\sqrt{s})}{m_R^2-s},
\end{equation}
where $a$, $b$ and $c$, $d$ are hadrons and the sum on $R$ runs over the number of resonances with mass $m_R$. The sum is restricted to the addition of resonances for a given spin $l$ and isospin $I$. 
The residue functions 
are given by
\begin{equation}
 g^2_{R\rightarrow ab}(\sqrt{s})=m_R\Gamma_{R\rightarrow ab}(\sqrt{s}),
\end{equation}
where $\sqrt{s}$ is the center of mass energy and $\Gamma_{R\rightarrow ab}(\sqrt{s})$ is
the energy dependent partial decay widths, i.e the total width times the branching ratio for the channel $R\rightarrow ab$ given as \cite{Chung:1995dx}
\begin{equation}\label{Eq:Width}
 \Gamma_{R\rightarrow ab}(\sqrt{s})=\Gamma^0_{R\rightarrow ab}\frac{m_R}{\sqrt{s}}\frac{ q_{ab}}{q_{ab0}}{\left(B^l(q_{ab},q_{ab0})\right)}^2.
\end{equation}
The momentum $q_{ab}$ is given as
\begin{equation}\label{Eq:commomentum}
 q_{ab}(\sqrt{s})=\frac{1}{2\sqrt{s}}\sqrt{\left(s-{(m_a+m_b)}^2\right)\left(s-{(m_a-m_b)}^2\right)},
\end{equation}
 where $m_a$ and $m_b$ being the mass of decaying hadrons $a$ and $b$.

In Eq.~($\ref{Eq:Width}$), $q_{ab0}=q_{ab}(m_R)$ is the resonance momentum at $\sqrt{s}=m_R$ and $\Gamma^0_R$ is the partial width of the pole at half maximum 
for the channel $R\rightarrow ab$. The $B^l(q_{ab},q_{ab0})$
are the Blatt-Weisskopf barrier factors which can be expressed in terms of momentum $q_{ab}$ and resonance momentum $q_{ab0}$
for the orbital angular momentum $l$. The detailed expression for $B^l(q_{ab},q_{ab0})$ can be found in Ref.~\cite{Chung:1995dx}.

\par

Furthermore, once one computes the K-matrix by providing the relevant masses and widths of resonances, the phase shift can be obtained using the relation:
\begin{equation}
  \delta_l^I=\tan^{-1} K(\sqrt{s}).
\end{equation}

Here we would like to note that a comparison between the empirical phase shifts of resonances 
and the K-matrix approach gives almost identical results for resonances like $\rho(770)$, $K^*(892)$, $N(1680)$, etc.

\subsection{Experimental Phase shifts}

\begin{figure*}
\centering
\includegraphics[width=0.45\textwidth]{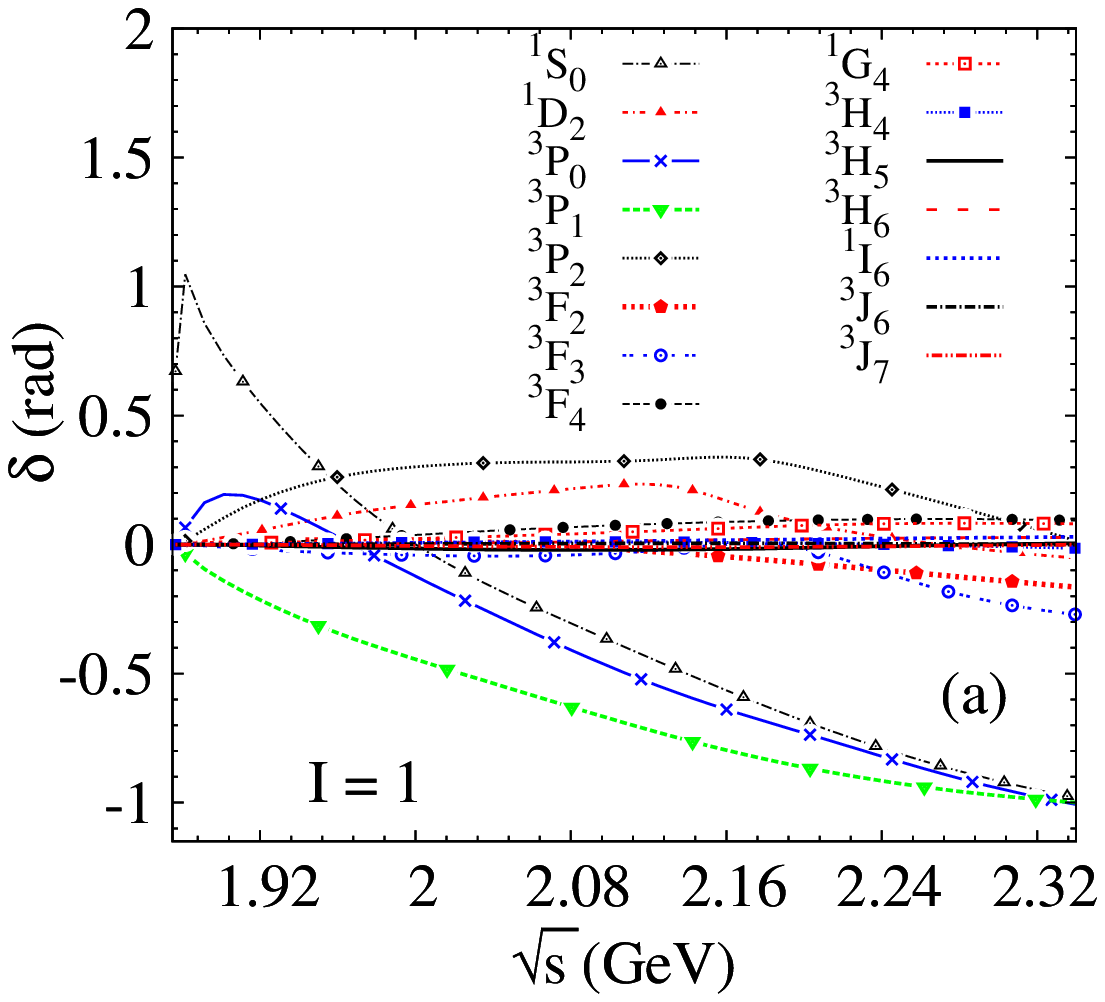}
\includegraphics[width=0.45\textwidth]{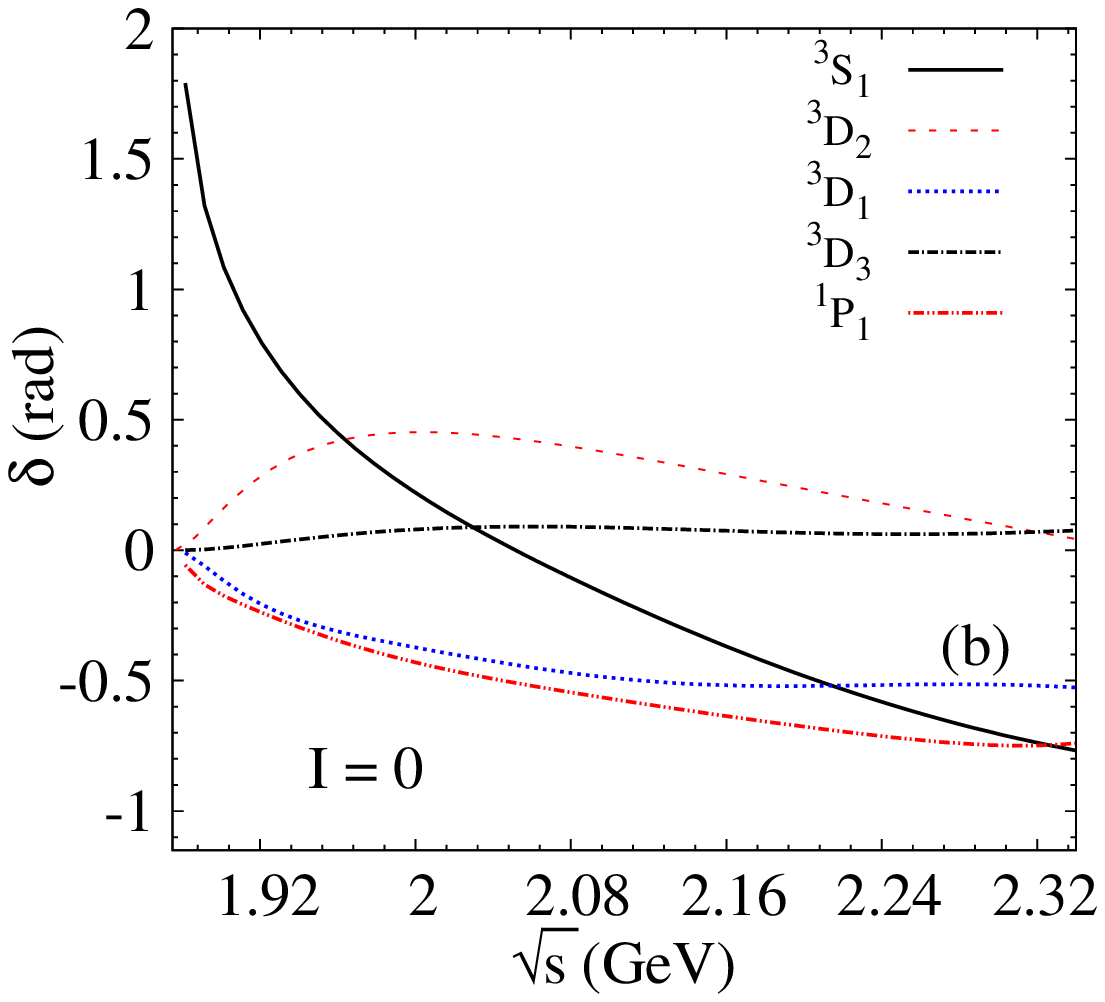}
\vspace{0.8cm}
 \caption{Energy dependence of NN scattering phase shifts taken
 from SAID partial-wave analysis \cite{Workman:2016ysf}.
 The notation to specify NN scattering channels is $^{2S+1}l_J$ where $l$, $S$, $J$ correspond to orbital, spin and total angular momentum respectively.}
\label{fig:NN_PhaseShift}
\end{figure*}

As mentioned earlier, for repulsive interactions and for interactions where the information about $m_R$ and $\Gamma_R$ are not available, the K-matrix formalism is not applicable and
we resort to extraction of phase shifts from experimental data. In our extraction of repulsive ($\pi N$, $KN$) and 
nucleon-nucleon ($NN$) interaction phase shifts, we use the data from the SM16 partial wave analysis \cite{Workman:2016ysf}. 
For the repulsive isotensor channel $\delta_0^2$ in the $\pi-\pi$ scattering, we use the data from Ref.~\cite{GarciaMartin:2011cn}. However, the S-matrix formalism elucidated here is only applicable for 
elastic scattering 
and the inelastic part that enters into the analysis by fitting to experimental data has to be removed. To get around this problem, we make an estimate of the contribution coming from the inelastic
part by first defining a generic $l$ dependent scattering amplitude $f_l(\sqrt {s})$,
\begin{equation}
f_l(\sqrt{s})=\frac{\eta_l e^{2i\delta_l}-1}{2i},
\end{equation}
where $\eta_l$ is the inelastic parameter. The elastic cross-section is given as, 
\begin{equation}\label{Eq:El}
\sigma_{\text{el}}=\frac{4\pi}{q^2}\sum_l (2l+1)\sin ^2 \delta_l,
\end{equation}
and the inelastic  cross-section is given by
\begin{equation}\label{Eq:InEl}
\sigma_{\text{inel}}=\frac{\pi}{q^2}\sum_l (2l+1)(1-\eta_l^2),
\end{equation}

\begin{figure*}
\begin{center}
\includegraphics[width=0.45\textwidth]{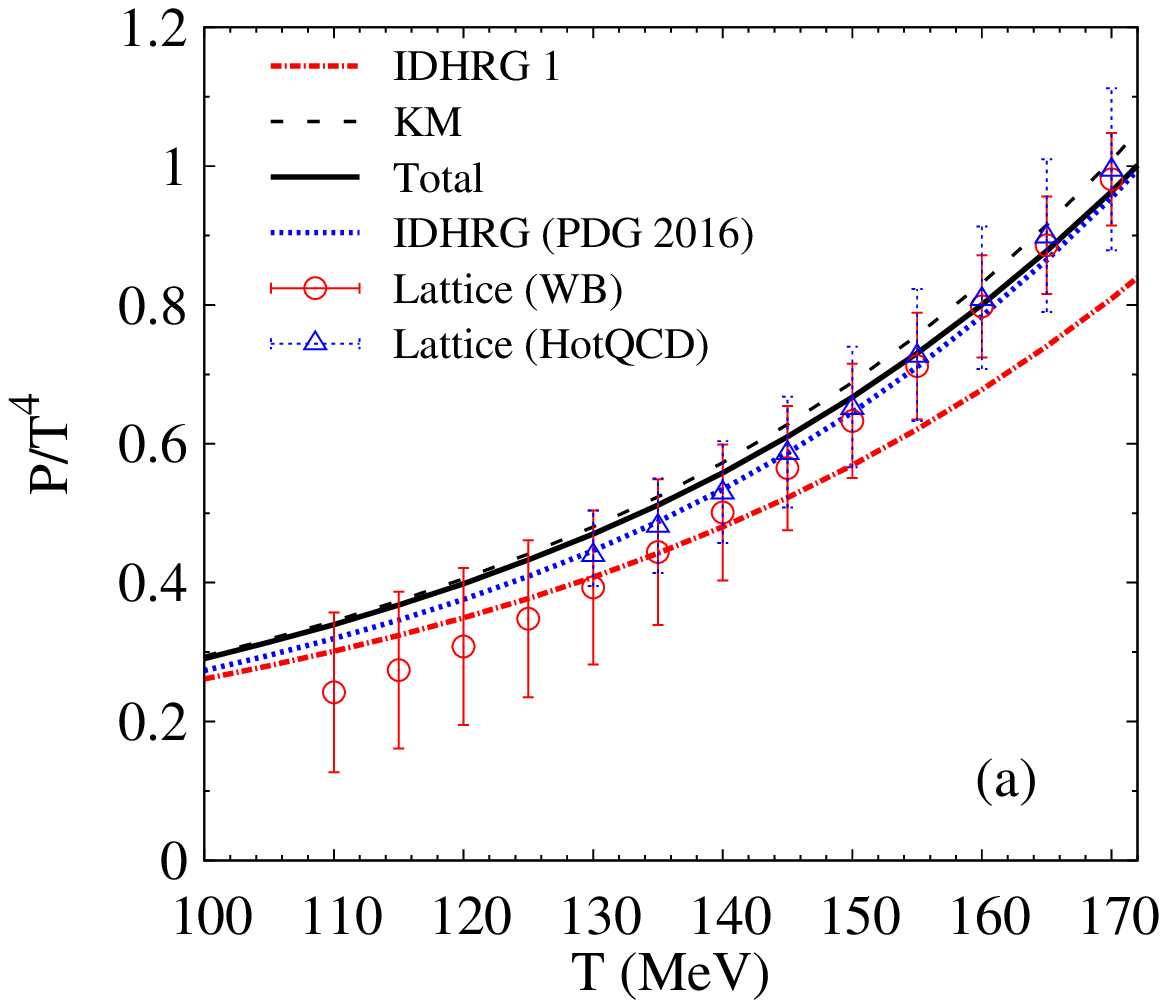}
\includegraphics[width=0.45\textwidth]{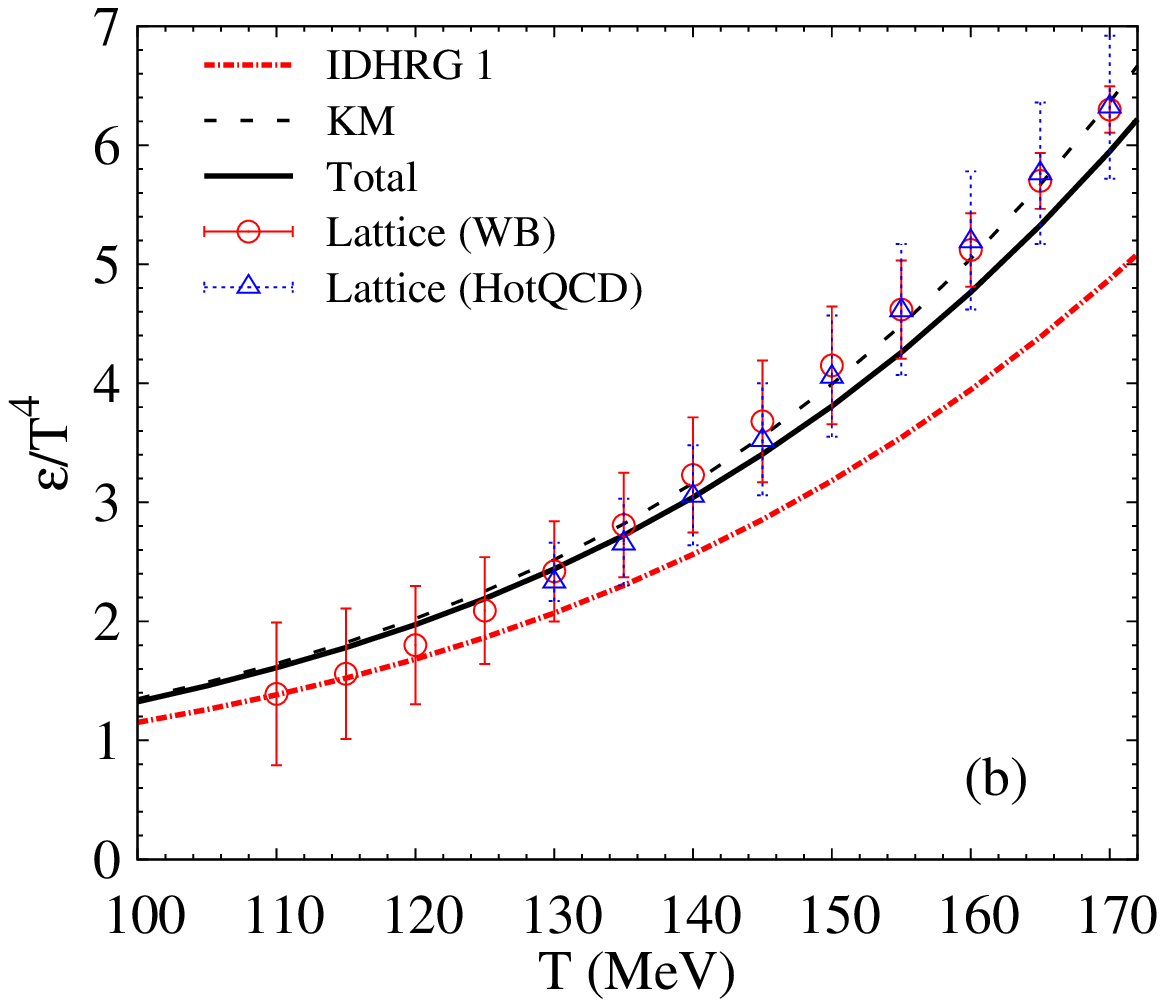}
\includegraphics[width=0.45\textwidth]{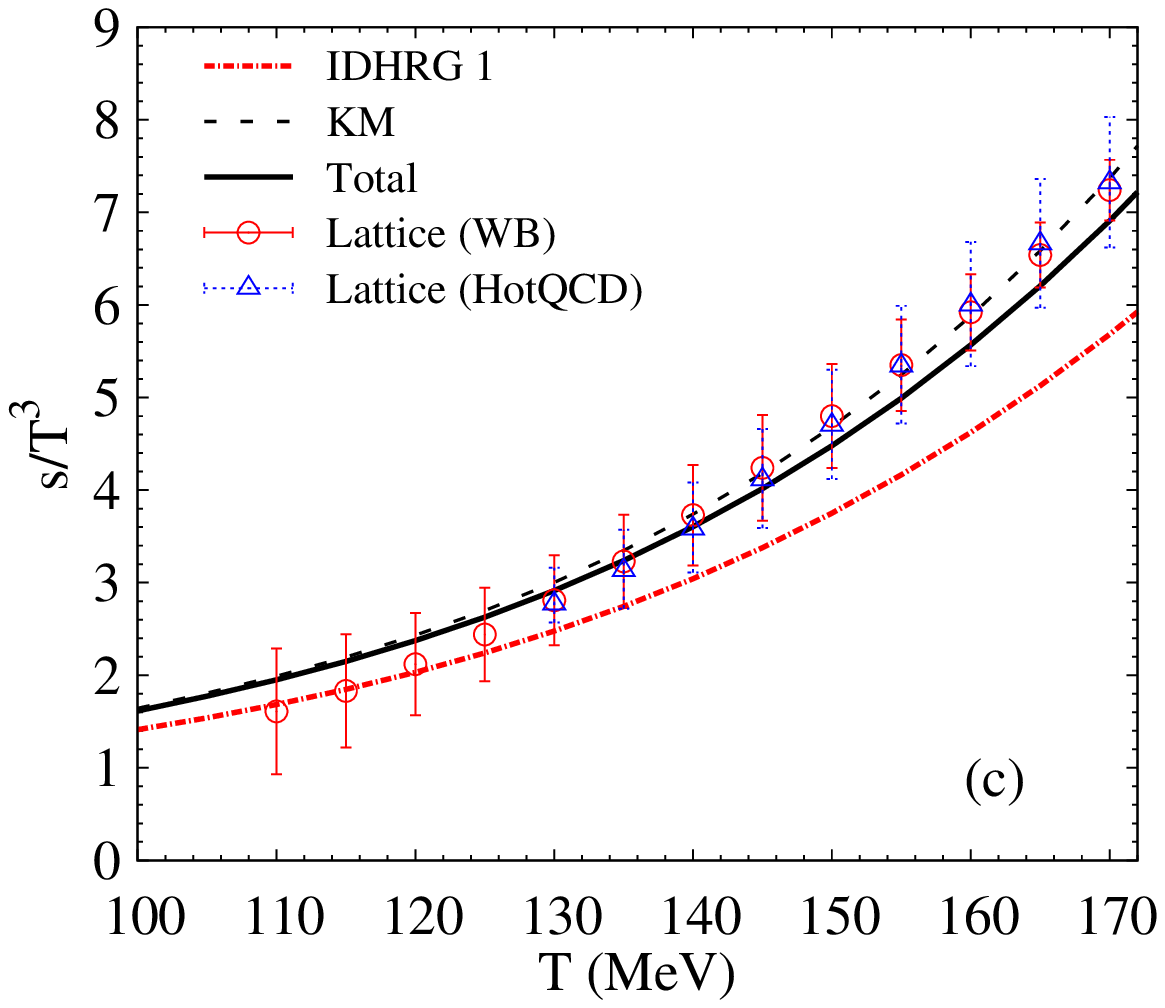}
\includegraphics[width=0.45\textwidth]{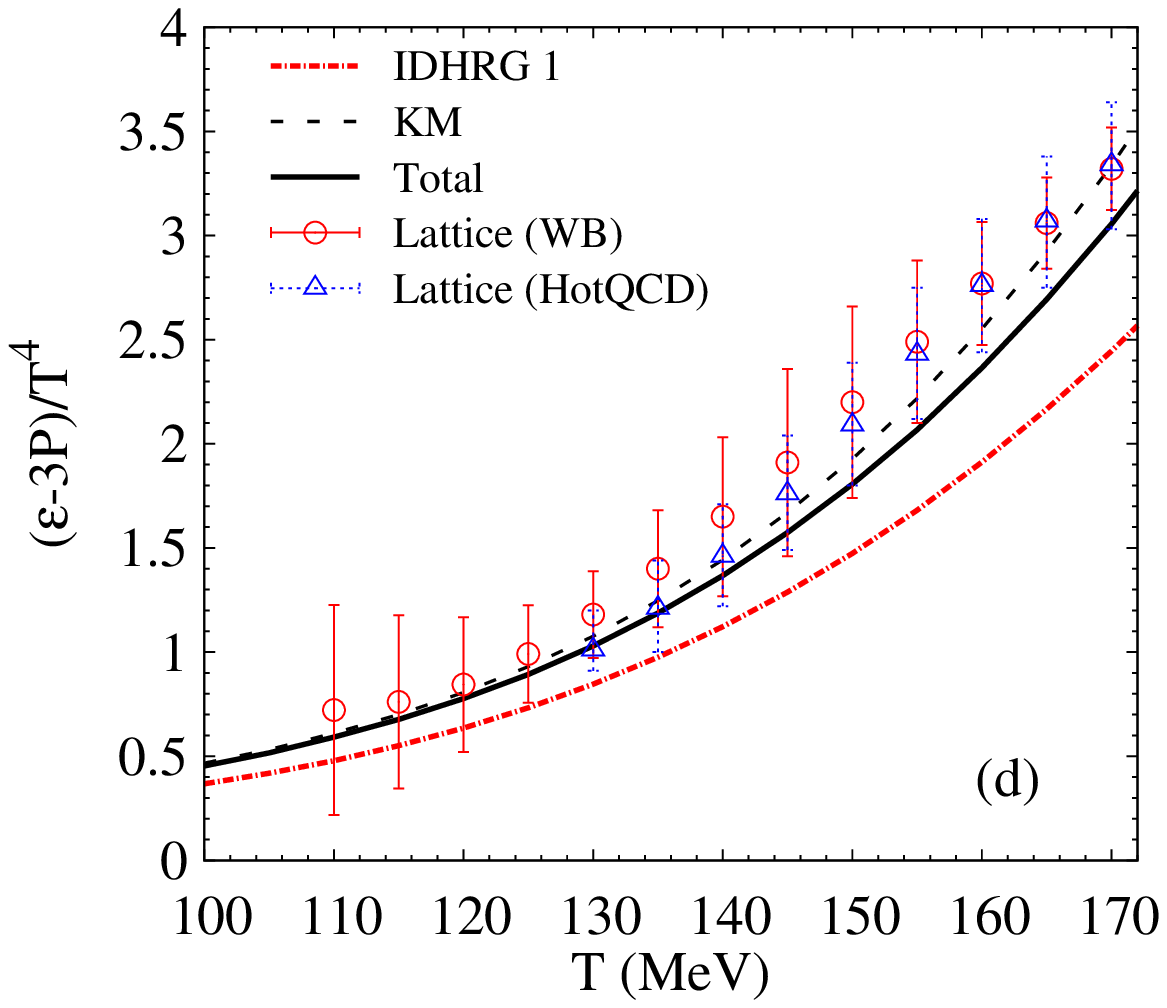}
\includegraphics[width=0.45\textwidth]{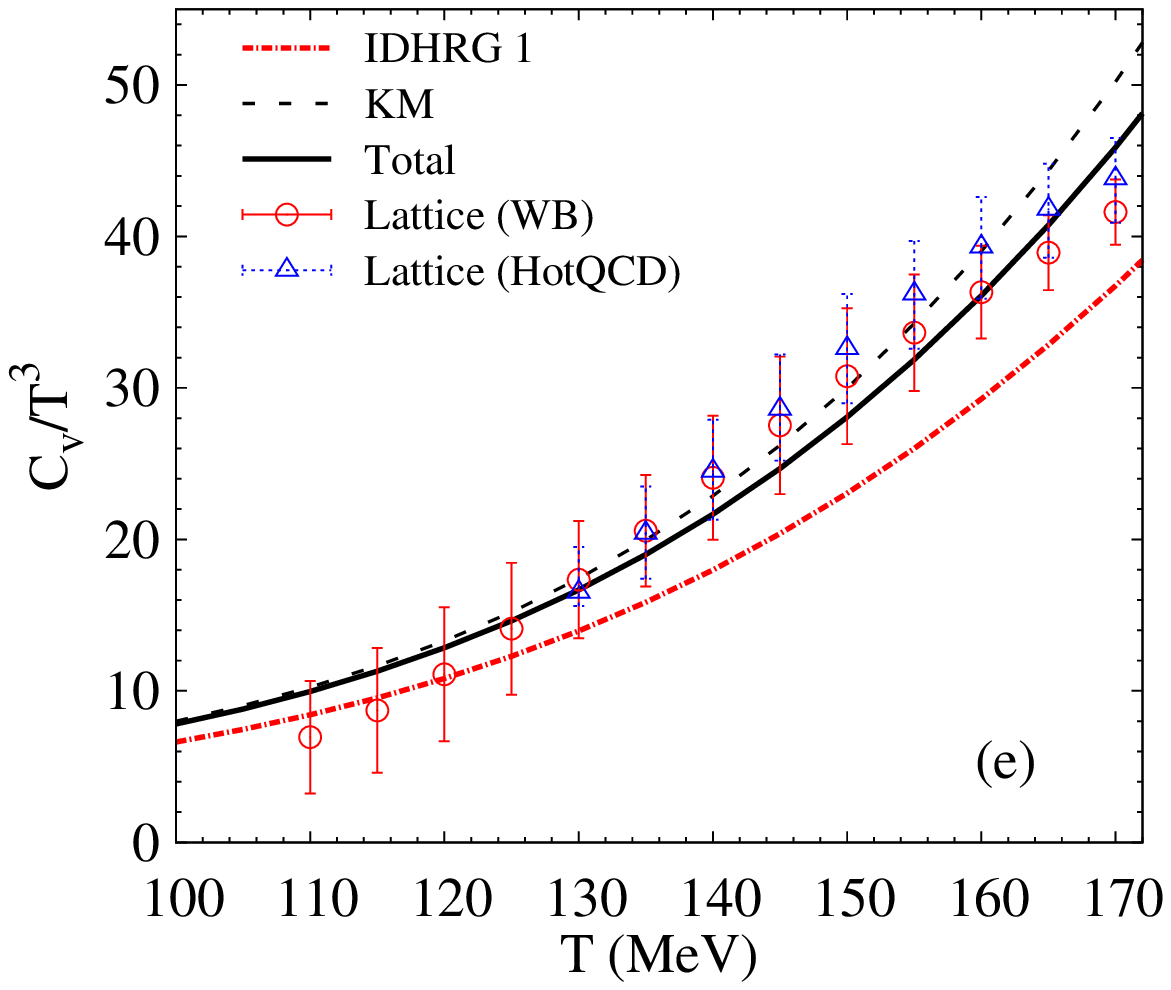}
\includegraphics[width=0.45\textwidth]{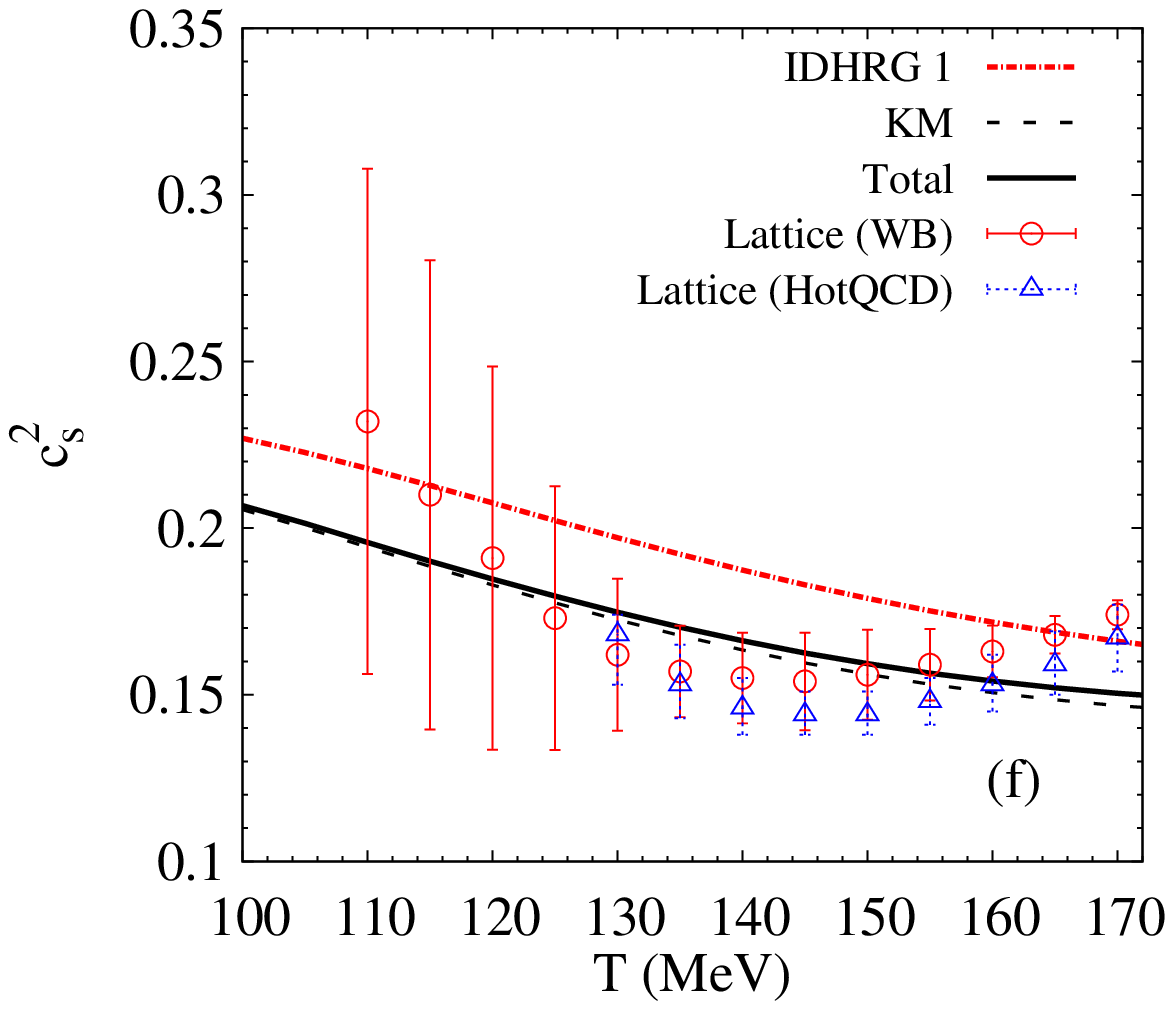}
\end{center}
\vspace{0.5cm}
 \caption{Temperature dependence of various thermodynamic quantities 
 ((a) $P/T^4$, (b) $\varepsilon/T^4$, (c) $s/T^3$, (d) $(\varepsilon-3P)/T^4$,
 (e) $C_V/T^3$ and (f) $c_S^2$) at zero chemical potential.
 Total contains both the attractive and repulsive interaction whereas KM contains only the 
 attractive part. IDHRG 1 corresponds to results of ideal HRG,
 with same number of particles as used in KM formalism.
 IDHRG (PDG 2016) in Fig.~\ref{fig:EOS}(a) corresponds to results of ideal HRG model for all the hadrons and
 resonances listed in PDG 2016 \cite{Patrignani:2016xqp}. Results are compared with lattice QCD data of
 Refs.~\cite{Borsanyi:2013bia} (WB) and \cite{Bazavov:2014pvz} (HotQCD).}
\label{fig:EOS}
\end{figure*}

where $q$ is center of mass momentum. The total cross section $\sigma$ is the sum of Eq.~(\ref{Eq:El}) and Eq.~(\ref{Eq:InEl}). We can approximate 
the contribution to the elastic part of the phase shift $\delta_{\text{el}}$ by the following expression
\begin{equation}\label{Eq:ElPart}
\delta_{\text{el}}\approx\sin^{-1}\left(\sqrt{\frac{\sigma_{\text{el}}}{\sigma}}\sin\delta\right),
\end{equation}
where $\delta$ is the total phase shift that is obtained from fit to experimental data \cite{Workman:2016ysf,Workman:2012hx,Hyslop:1992cs}.\par

\subsubsection{N-N interactions}
For the nucleon-nucleon ($NN$) interaction we have included the phase shifts for $l\leq 7$ in both $I=0$ and $I=1$ isospin channels. 
Combinations of $l$, $S$ and $I$ are chosen so that the total wave function for $NN$ interaction is anti-symmetric as dictated by Pauli's principle. 
We have restricted the range of energies 
up to the pion ($\pi$) production threshold. Beyond this threshold, the contribution from the inelastic channels become dominant and the present formalism fails to disentangle the contribution from the elastic and inelastic part. 
However, below this threshold where the contribution from inelasticities are sub-dominant, we can extract the contribution from the elastic part using the approximation Eq.~(\ref{Eq:ElPart}). In order 
to use Eq.~(\ref{Eq:ElPart}) we need a parametrization of the cross-section as a function of energy which in the present study are used from Ref.~\cite{Cugnon:1996kh}
\begin{eqnarray}\label{Eq:sigmatot}
 \sigma^{NN} (~\text{mb})&=&23.5+1000(p_{\text{lab}}-0.7)^4,~p_{\text{lab}}<0.8~\text{GeV}\nonumber\\
  &=& 23.5+\frac{24.6}{1+\exp{\left(-\frac{p_{\text{lab}}-1.2}{0.10} \right)}},\nonumber\\
 && 0.8<p_{\text{lab}}<5 ~\text{GeV}\nonumber\\
 &=& 41+60(p_{\text{lab}}-0.9)\exp(-1.2 p_{\text{lab}}),\nonumber\\
 && 1.5<p_{\text{lab}}<5 ~\text{GeV},
\end{eqnarray}
where $p_{\text{lab}}$ is the laboratory momentum. Similarly the elastic cross-section $\sigma_{\text{el}}$ can be parametrized as
\begin{eqnarray}\label{Eq:sigmael}
 \sigma_{\text{el}}^{NN} (~\text{mb})&=&23.5+1000(p_{\text{lab}}-0.7)^4,~p_{\text{lab}}<0.8 ~\text{GeV}\nonumber\\
 &=& \frac{1250}{p_{\text{lab}}+50}-4{(p_{\text{lab}}-1.3)}^2,\nonumber\\
 && 0.8<p_{\text{lab}}<2 ~\text{GeV}\nonumber\\
 &=&\frac{77}{p_{\text{lab}}+1.5}, p_{\text{lab}}>2 ~\text{GeV.}\nonumber\\
\end{eqnarray}
By comparing Eq.~(\ref{Eq:sigmatot}) and Eq.~(\ref{Eq:sigmael}), we can see that the contribution 
from inelastic processes is small below $p_{\text{lab}}<0.8$~GeV and increases
further with $p_{\text{lab}}$. In Fig.~\ref{fig:NN_PhaseShift} we have plotted the experimental NN phase 
shifts from the SAID partial-wave analysis \cite{Workman:2016ysf}
as a function of center of mass energy ($\sqrt{s}$).
Dominant contribution 
comes from lower $l$ values e.g. the $^{1}S_0$ phase shift which peaks at lower $\sqrt{s}$ and then falls sharply or the rapidly falling and largely repulsive $^{3}S_1$ phase shift. 
An interesting case to observe are the triplet P-waves which can have $J=0$, 1, 2 corresponding to phase shifts $^{3}P_0$, $^{3}P_1$, $^{3}P_2$. The behavior of the phase shifts are quite 
different in the above three channels, from zero crossing to purely repulsive and purely attractive case as seen in Fig.~\ref{fig:NN_PhaseShift}. 
This could be attributed to the spin-orbit 
coupling which 
splits them in to the triplet states having different behavior depending on the sign and strength of the coupling. However, most of the phase shifts become negative at higher $\sqrt{s}$
signifying the hard core nature of NN interaction. 
We would like to note that for NN interaction, the contribution from bound
states e.g., the $^3S_1$ channel which forms deuteron
at threshold is not taken into account in
Eq.~(\ref{Eq:IntPressure}).\par

\begin{figure*}
\begin{center}
\includegraphics[width=0.45\textwidth]{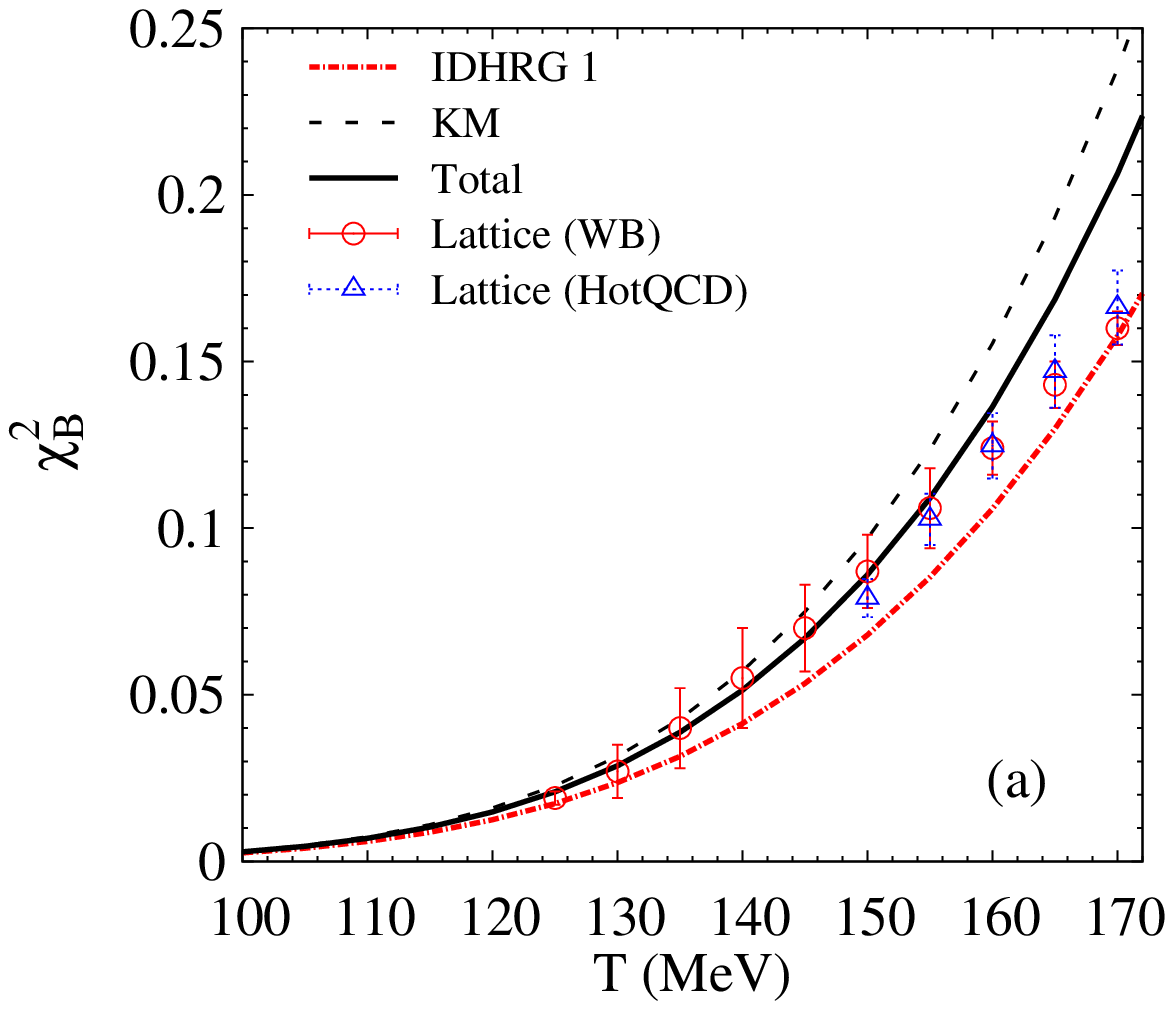}
\includegraphics[width=0.45\textwidth]{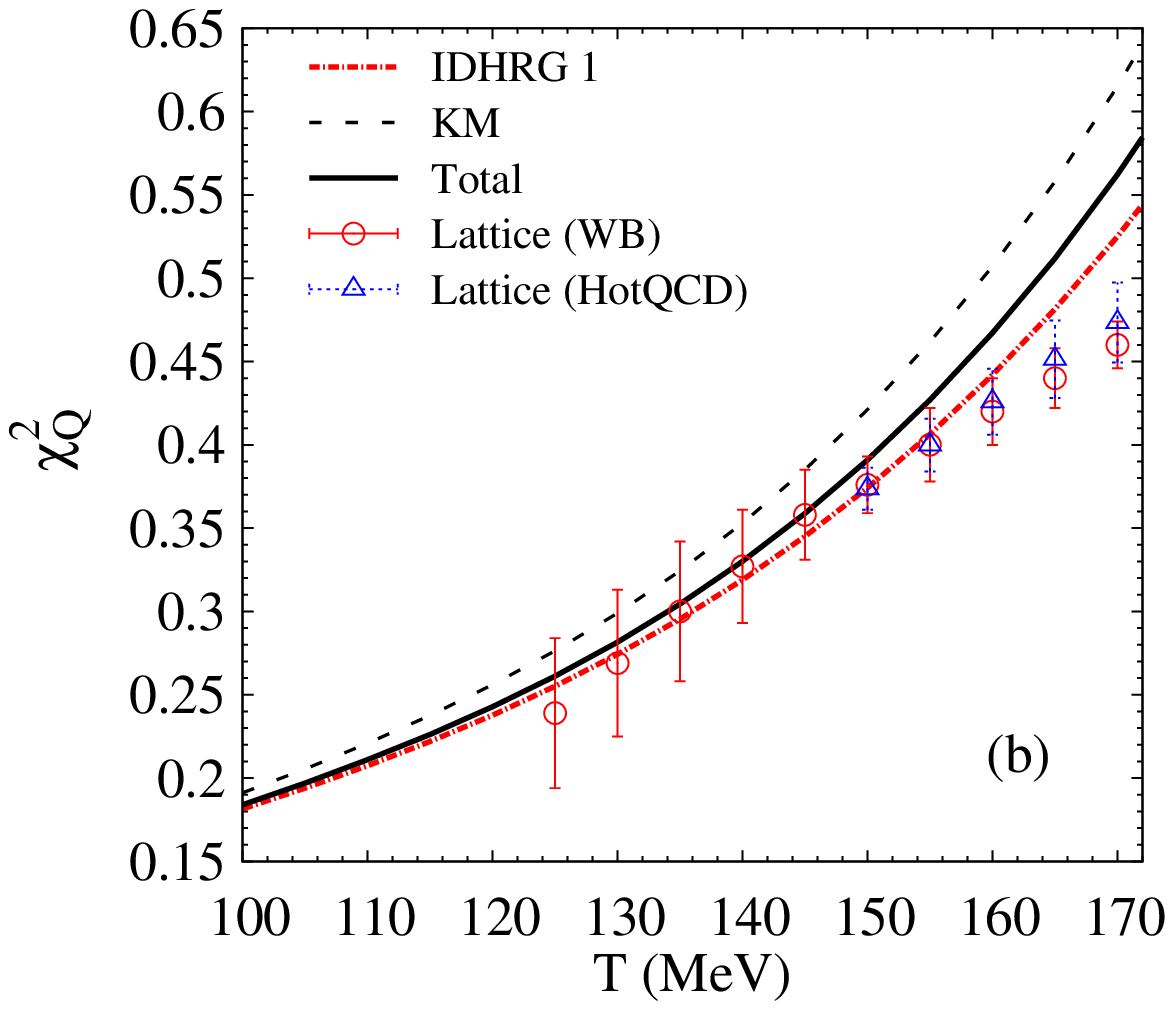}
\includegraphics[width=0.45\textwidth]{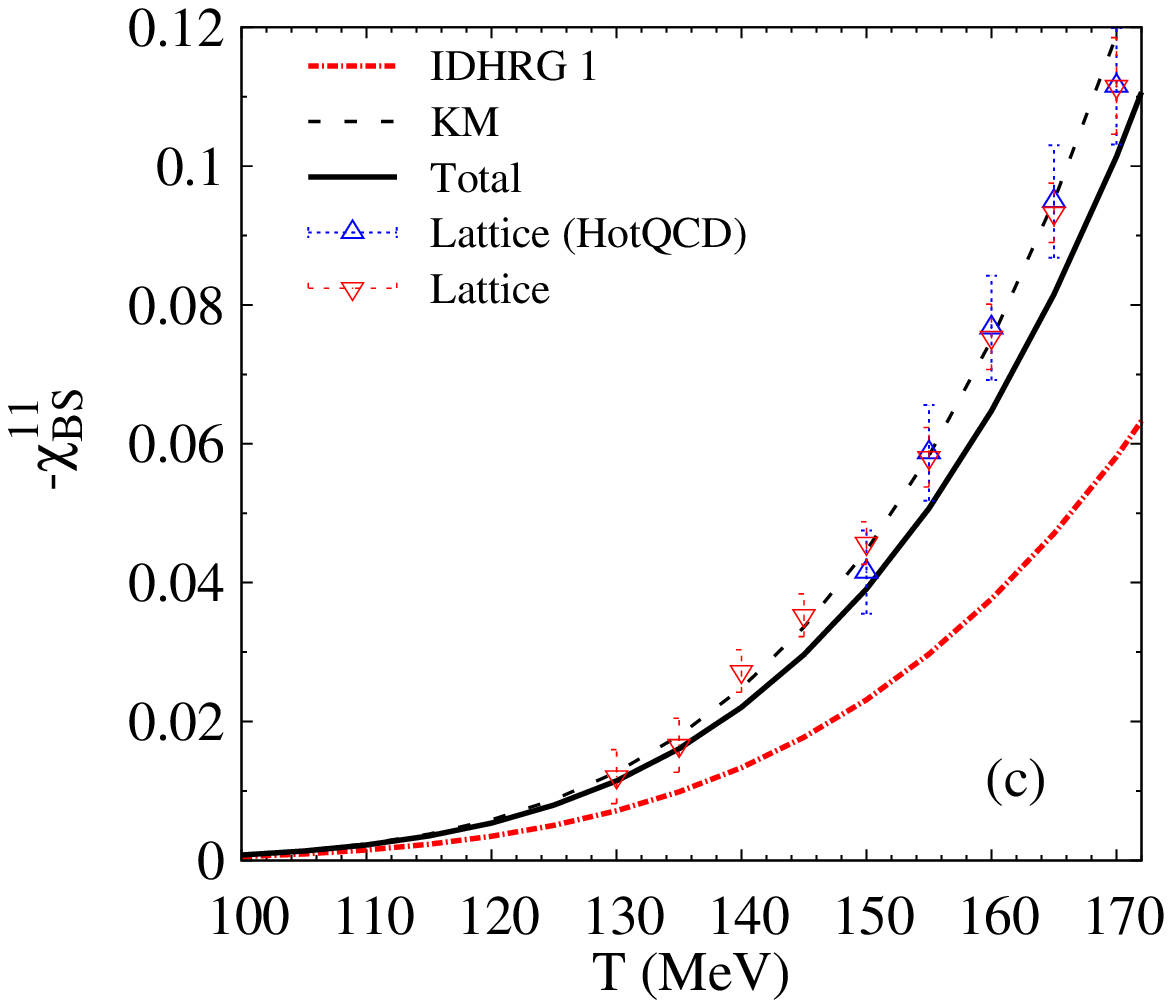}
\includegraphics[width=0.45\textwidth]{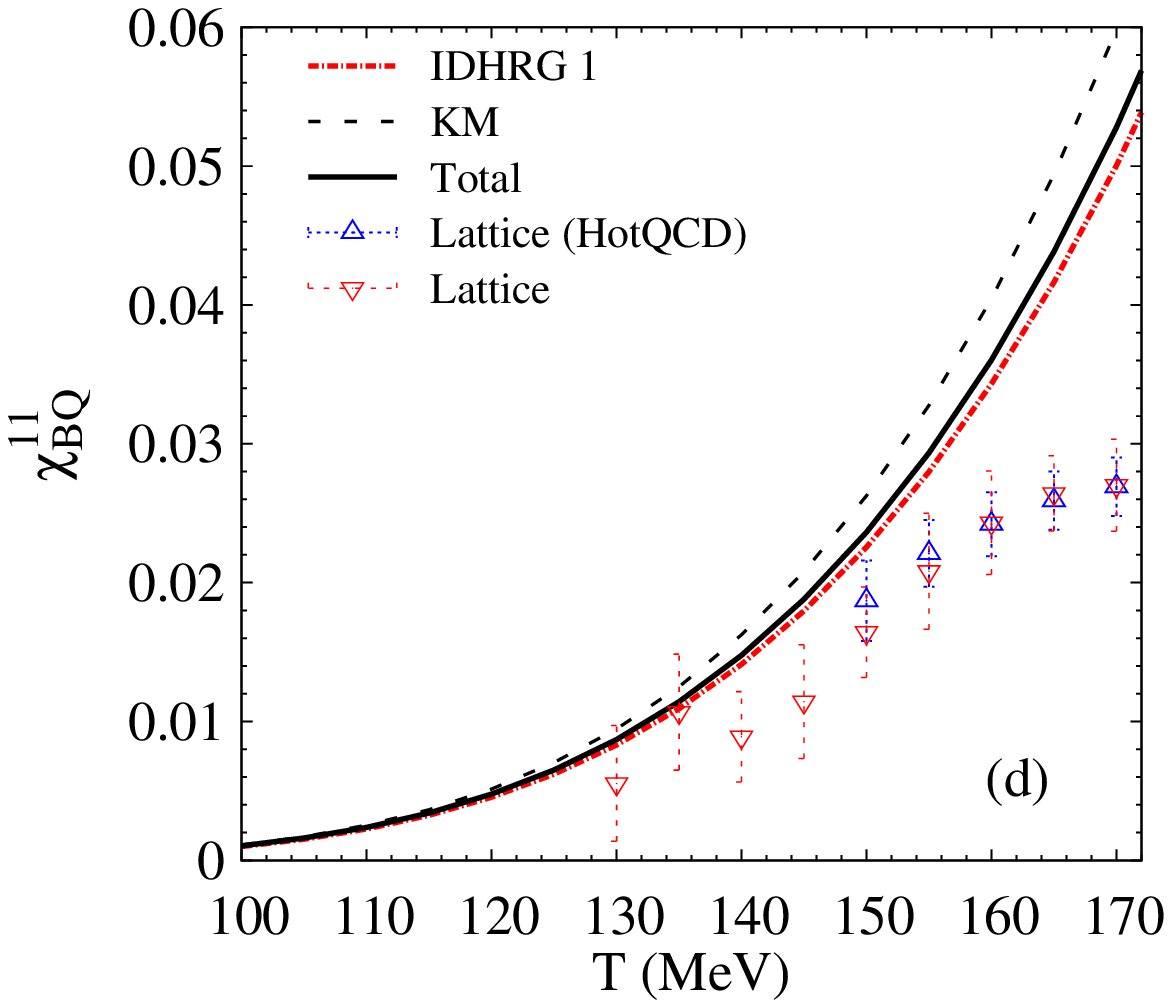}
\end{center}
\vspace{0.5cm}
 \caption{Temperature dependence of second order susceptibilities
((a) $\chi^2_B$, (b) $\chi^2_Q$, (c) $\chi^{11}_{BS}$ and (d)  $\chi^{11}_{BQ}$) at zero chemical potential.
 Total contains both the attractive and repulsive interaction whereas KM contains only the 
 attractive part. IDHRG 1 corresponds to results of ideal HRG,
 with same number of particles as used in KM formalism. Results are compared with lattice QCD data of Refs.~\cite{Borsanyi:2011sw} (WB),
~\cite{Bazavov:2012jq} (HotQCD) and~\cite{Bellwied:2015lba} (Lattice).}
\label{fig:chidiagonal}
\end{figure*}

\subsubsection{$\pi$-N interactions}
For pion-nucleon ($\pi N$) interaction we have included only those phase shifts \cite{Workman:2012hx} which are purely
repulsive and the attractive part are from the K-matrix parametrization. Here, 
we have restricted the energies to the eta ($\eta$) production threshold and the 
cross-sections are parametrized from Ref.~\cite{Cugnon:1996kh} using isobar model as
\begin{eqnarray}\label{Eq:sigmaInel}
 \sigma^{\pi N} (~\text{mb})&=&\frac{326.5}{1+4{\left(\frac{\sqrt{s}-1.215}{0.110}\right)}^2}\frac{q^3}{q^3+{(0.18)}^3},\nonumber\\
\end{eqnarray}
where $q$ is the center of mass momentum. In the range of momenta $0.5$~GeV$<p_{\text{lab}}<1.5$~GeV, the inelastic channel $\pi N\rightarrow \pi\pi N$ is the
most dominant whose cross section can be parametrized as
\begin{equation}
 \sigma_{\text{inel}}^{\pi N} (~\text{mb})=74{(p_{\text{lab}}-0.555)}^2 p_{\text{lab}}^{-4.04}~\text{GeV.}
\end{equation}

The dominant repulsive contribution in the $\pi N$ interaction comes from the $S_{31}$ ($l_{2I,2J}$) phase shift 
corresponding to $\Delta (1620)$ resonance. We would like to stress 
here that in our previous study of resonances in Ref.~\cite{Dash:2018can}, using K-matrix formalism, 
resonances such as
$\Delta (1620)$, $\Delta (1910)$, $\Delta (1930)$ and $N (1720)$ were included in the attractive part
of the S-matrix, via their masses and partial decay widths (i.e. branching fraction times the total width) for a resonance $R$ interacting through the process $ab\rightarrow R\rightarrow ab$, 
where $a$ and $b$ are the corresponding hadrons. However, a comparison to experimental phase shifts
through the factor $\partial\delta^l(\epsilon)/\partial\epsilon$ has rendered that, it is negative below the $\eta$ production threshold. Thus, such resonances are included in the repulsive part by fitting to 
experimental phase shifts.\par

\subsubsection{K-N interactions}
For the $KN$ interaction, the dominant repulsive contribution comes from the $S_{11}$ ($l_{I,2J}$) phase shift containing 
the $\Sigma(1660)$ resonance. 
Similar to the case of $\pi N$, $\Sigma$ resonances like 
$\Sigma(1660)$, $\Sigma(1750)$ and $\Sigma(1915)$; 
$\Lambda$ resonances like $\Lambda(1520)$, $\Lambda(1600)$ and $\Lambda(1690)$ 
were considered 
attractive in \cite{Dash:2018can}, but here we include them in the repulsive part, since  $\partial\delta^l(\epsilon)/\partial\epsilon$
is negative below the inelastic production threshold \cite{Hyslop:1992cs}. 
The cross sections are parametrized from Ref.~\cite{STIN:1989} as
\begin{eqnarray}
\sigma^{KN} (~\text{mb})&=&23.91+17.0\exp\left(-\frac{{(p_{\text{lab}}-10)}^2}{0.12}\right),\nonumber\\
&&~p_{\text{lab}}<2.5~\text{GeV and}\nonumber\\
 \sigma_{\text{el}}^{KN} (~\text{mb})&=&172.38\exp{\left( -2.0(p_{\text{lab}}+0.1)\right)},\nonumber\\
&&~p_{\text{lab}}<0.7~\text{GeV.}
\end{eqnarray}

\subsubsection{$\pi$-$\pi$ interactions}
For the pion-pion ($\pi\pi$) interaction we have included the dominant repulsive phase shift from Ref.~\cite{GarciaMartin:2011cn},
in the isotensor channel $\delta_0^2$, as does in previous studies
\cite{Venugopalan:1992hy,Broniowski:2015oha}. 
This phase shift is known to cancel the isoscalar channel $\delta_0^0$ containing the broad $f_0(500)$ ($\sigma$ meson). The relevant energies have been restricted to pion production 
threshold.

\section{\label{sec:Result}Result}
In Fig.~\ref{fig:EOS}, we show the temperature variation at zero chemical potential 
for various thermodynamic quantities such as scaled pressure, energy density, 
entropy density, speed of sound and the specific heat capacity at constant volume. 
Results of attractive K-matrix (KM) based HRG model from Ref.~\cite{Dash:2018can} are compared with the 
total contribution (Total), which contains both attractive and repulsive channels done in the present
work. In Ref.~\cite{Dash:2018can} it was found that the effect of attractive interaction through KM approach increases 
the value of all thermodynamic observables compared to the ideal HRG result (IDHRG 1). 
It must be noted that the K-matrix
formalism includes only those resonances which have two body decay mode and only these resonances were included in IDHRG 1. 
We observe that the effect of repulsive interactions cancels some of the contributions
from attractive channels, thereby slightly lowering the net result for Total relative to KM for
the observables studied here. A second comparison with ideal HRG Fig.~\ref{fig:EOS}(a), that considers all the 
confirmed hadrons and resonances consisting of up, down, 
and strange flavor valence quarks listed in the PDG 2016 
Review \cite{Patrignani:2016xqp} (IDHRG (PDG 2016)), shows a better agreement with lattice data.
However, it is worth mentioning here that the agreement of 'IDHRG (PDG 2016)'
with the LQCD data is because of the increase in the number of degeneracies and not due 
to some inherent interaction that is naturally present in the system revealed within the S-matrix formalism. On the
whole, we conclude that the effect of repulsive channels suppress the bulk variables
studied here, compared to K-matrix (KM) approach and are  shown in Fig.~\ref{fig:EOS}.

Similar to thermodynamic observables, it was found 
in Ref.~\cite{Dash:2018can} that the K-matrix formalism leads to an increment in the values of 
diagonal and off diagonal susceptibilities compared to ideal HRG result.
The effect of repulsive interactions are most prominent when we calculate these second order 
diagonal and off diagonal susceptibilities.
Results for $\chi^2_B$, $\chi^2_Q$,  $\chi^{11}_{BS}$ and
$\chi_{BQ}^{11}$, ($B,Q$ stand for baryon and electric charge respectively. Definition of
susceptibilities can be found in Ref.~\cite{Dash:2018can}) shown in Fig.~\ref{fig:chidiagonal}, agree better 
with the LQCD data, in the case when both attraction and repulsion are taken into
account than in the K-matrix formalism. The effect of repulsion is mostly visible in 
the baryonic sector. For example, we have checked for $\chi^2_B$, the contribution from repulsive
interaction has the following order $\pi N>KN>NN$. Although, we find that many channels are 
repulsive in $NN$ interaction
than in $\pi N$ interaction, the effect of repulsion on observables like $\chi^2_B$ is more 
from $\pi N$ interaction. This is 
because the effect of repulsion in elastic $\pi N$ interaction is dominant in the energy 
ranges $1.07$~GeV$<\sqrt{s}<1.67$~GeV,
while for elastic $NN$ interaction is in between $1.88$~GeV$<\sqrt{s}<2.34$~GeV. This fact is
reflected when we compute thermodynamic
observables in the relevant temperature ranges.
We have also checked for the remaining second order diagonal
and off diagonal susceptibilities, and the difference between Total and the K-matrix formalism for $\chi^2_S$ is small and for $\chi^{11}_{QS}$
is negligible.

\par

\begin{figure*}
\begin{center}
\includegraphics[width=0.45\textwidth]{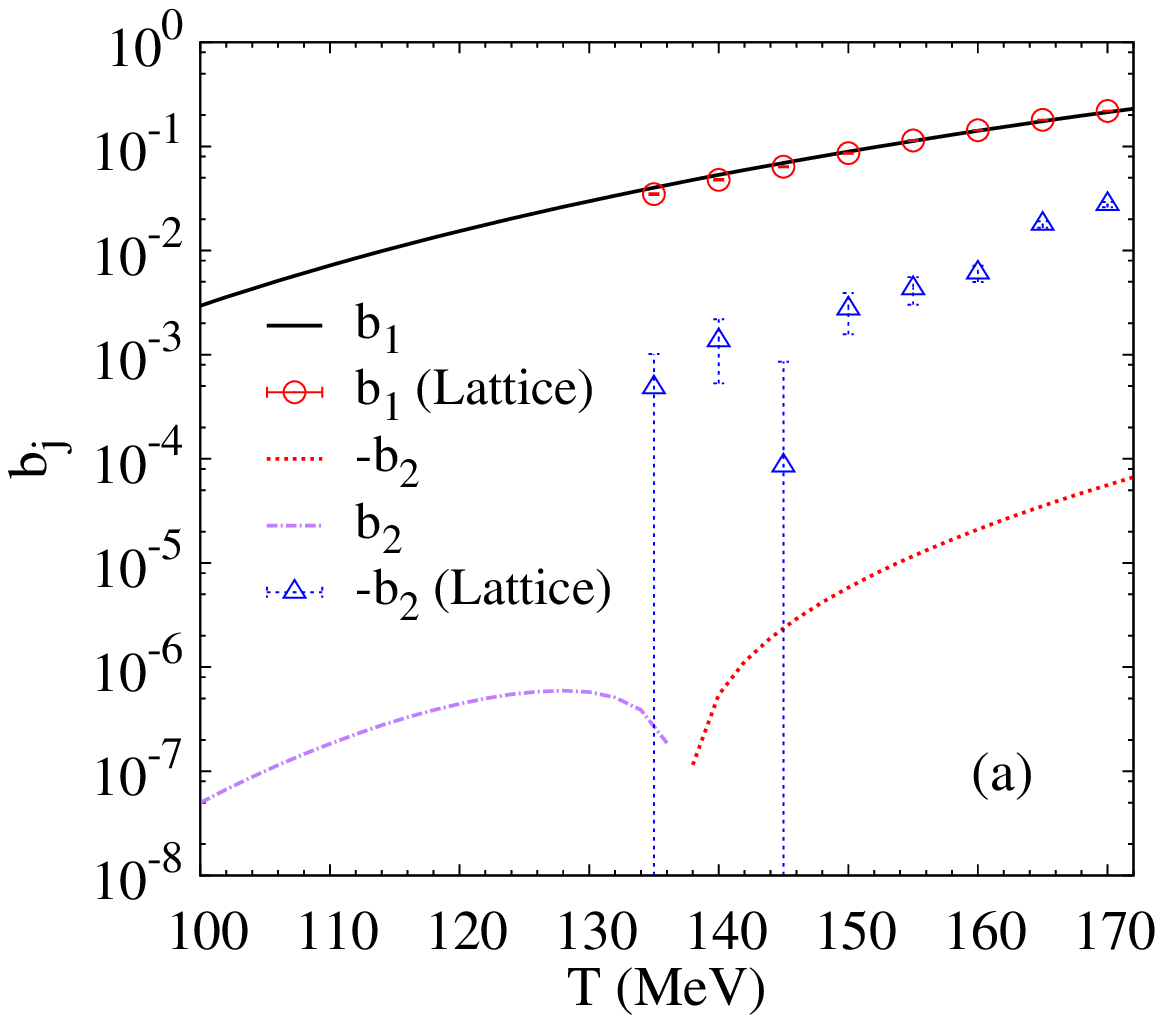}
\includegraphics[width=0.45\textwidth]{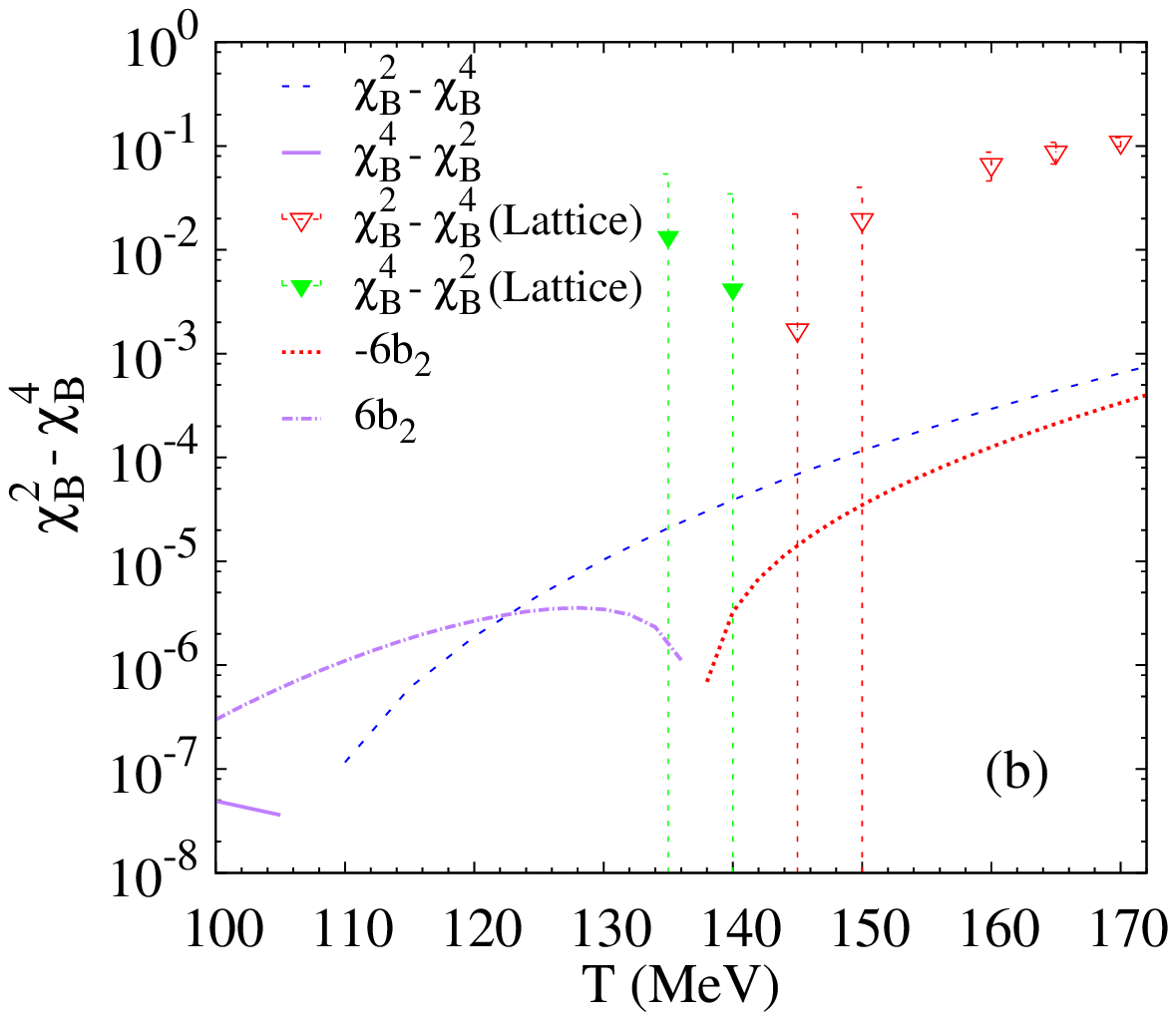}
\end{center}
\vspace{0.5cm}
 \caption{Left panel shows the variation of Fourier coefficients $b_j(T)$ with temperature computed using S-matrix formalism compared to
 the lattice results from Ref.~\cite{Vovchenko:2017xad}. Open symbols (circles and triangles) denote the result from lattice QCD.
 Solid (black) line denotes the result of Fourier coefficient $b_1(T)$. 
 Dot dashed line with purple and dot lines with red color represent
 the positive and negative parts of the Fourier coefficient $b_2(T)$ respectively.
 Right panel shows the variation of $\chi^2_B - \chi^4_B$ with temperature at zero chemical
 potential. Non zero value comes mainly due the $NN$ interaction which is shown
 by the dashed blue and solid purple lines, denoting the positive and negative parts assuming $FD$ statistics in the ideal part.
 Dot dashed line with purple and dot line with red color represent
 the positive and negative parts of the Fourier coefficient $6b_2(T)$ respectively (see text).
 Result is compared with lattice QCD data of Ref.~\cite{Bellwied:2015lba}
 (Lattice) with open red and solid green symbols denoting the
 positive and negative parts.}
\label{fig:SeondOrder}
\end{figure*}

Lattice observables like fluctuations and correlations of conserved charges at finite net baryon density is expected to
be sensitive to the modeling of baryonic interactions. However, lattice calculation at finite $\mu_B$ is not possible because 
of the sign problem. Methods like Taylor series expansion and analytic continuation from imaginary $\mu_B$ has been devised to
get around this problem \cite{Allton:2002zi,Basak:2009uv,Gavai:2008zr,deForcrand:2002hgr,DElia:2002tig,Wu:2006su,deForcrand:2008vr,
DElia:2009pdy,Philipsen:2014rpa,Cuteri:2015qkq,DElia:2016jqh,Alba:2017mqu}. 
Following Ref.~\cite{Vovchenko:2017xad}, we directly compare the predictions of S-matrix formalism to lattice data at
imaginary chemical potential instead of performing analytic continuation to real chemical potential. Since, the QCD pressure
is an even function of real $\mu_B$, the first order net baryon susceptibility assuming Maxwell-Boltzmann (MB) statistics, can be written as \cite{Vovchenko:2017xad},
\begin{equation}
 \chi^1_B=\sum_{j=1}^{\infty} b_j \sinh(j\mu_B/T),
\end{equation}
where $b_j$ contains the information from different baryonic sectors. Using analytic continuation one can convert
the above sum to a Fourier series expansion where the Fourier coefficients are given as,
\begin{equation}
 b_j(T)=\frac{2}{\pi}\int_0^{\pi}dx~\text{Im}\left[\chi_B^1(T,i\mu_B)\right]\sin (jx),
\end{equation}
where $x=\mu_B/T$.\par

The results of two leading order Fourier coefficients $b_1(T)$, $b_2(T)$ computed using S-matrix formalism are compared to lattice
results and are shown in Fig.~\ref{fig:SeondOrder}(a). We find a very good agreement between the coefficient $b_1(T)$ and lattice QCD results using
imaginary chemical potential. Moreover, we found that $b_2(T)$ is quite small compared to the lattice
results which is due to the inclusion of only $NN$ interaction and not other baryon-baryon interaction. However, we found that
$b_2(T)$ is positive for $T<135$ MeV and negative above this temperature contrary to the results of
Ref.~\cite{Vovchenko:2017xad} which is negative throughout the temperature range. This can be understood from the isospin 
weighted sum of phase-shifts of $NN$ interaction which is positive for small $\sqrt{s}$ and falls 
rapidly at large $\sqrt{s}$ showing the hard core nature of $NN$ interaction at short distances.

The contribution from interaction can be explored further by considering 
certain combinations of diagonal and off diagonal susceptibilities which are identically
zero for non-interacting or ideal HRG but not for non-interacting gas of quarks and 
gluons \cite{Bazavov:2013dta}. The quantity $\chi^2_B-\chi^4_B=0$, for a hadron gas 
which has baryon number $\pm1$, but not for non-interacting QGP for which  $\chi^2_B-\chi^4_B>0$, since
all quarks carry a baryon number of $\pm1/3$. However for an interacting gas, the 
inclusion of $NN$ interaction which carries a net baryon number $\pm 2$ might give us
a non zero result. It is particularly instructive to note that this observable is related to Fourier coefficients
such that $\chi^2_B-\chi^4_B=-\sum_{j=2}^{j=\infty}j(j^2-1)b_j(T)$ assuming MB statistics. In
our case, since the contribution comes only from the term $j=2$ for $NN$ interaction, we have 
$\chi^2_B-\chi^4_B=-6b_2(T)$. We compare this with the S-matrix formalism where the ideal part is 
computed assuming Fermi-Dirac (FD) statistics.
This is shown in Fig.~\ref{fig:SeondOrder}(b), and the result shows
that $-6b_2(T)$ changes sign in accordance to the discussion in the previous paragraph. However, we find that
the influence of statistics (FD) in S-matrix formalism leads to a increase in the value of observable $\chi^2_B-\chi^4_B$ and shifts
the change in sign to a lower temperature. The above observation is in agreement with lattice data which also shows a similar change in sign
when moving from lower to higher temperature.
For temperatures $T>110$ MeV we find that $\chi^2_B-\chi^4_B>0$ again indicating the hard core nature of $NN$ 
interaction. In Ref.~\cite{Huovinen:2017ogf,Vovchenko:2016rkn} the same increasing 
trend of $\chi^2_B-\chi^4_B$ with temperature was also found using repulsive mean 
field in a multi-component hadron gas and excluded volume approach. Our results 
using the S-matrix formalism validate the previous results. 
Moreover, one should note that the effect of including 
only $NN$ interaction is rather small compared to the results obtained by Ref.~\cite{Huovinen:2017ogf,Vovchenko:2016rkn}
which can be improved upon adding other baryon-baryon interaction in the partition function. 
However, we do not have information about the experimental phase shifts of other baryon-baryon interactions and one 
has to take recluse of chiral effective theory \cite{Polinder:2006zh,Haidenbauer:2013oca} or other such methods which 
is left as a future work. Other observables 
like $v_1=\chi^{31}_{BS}-\chi^{11}_{BS}$ and $v_2=1/3\left(\chi^2_{S}-\chi^4_{S}\right)-2\chi_{BS}^{13}-4\chi_{BS}^{22}-2\chi_{BS}^{31}$, 
\cite{Bazavov:2013dta} are trivially zero in our analysis since 
we do not have the information about interactions (phase shifts) among baryons which
have $|B|>1$ and $|S|=1$ or vice-versa.\par

\begin{figure}
\begin{center}
\includegraphics[width=0.45\textwidth]{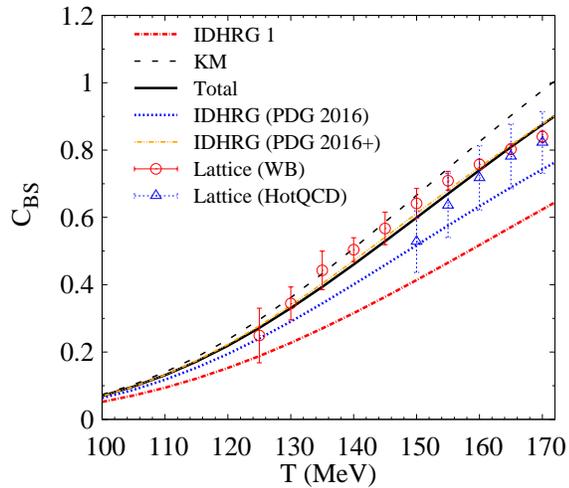}
\end{center}
\vspace{0.5cm}
 \caption{The temperature dependence of $C_{BS}$ at zero chemical potential calculated in
 the current work (Total). IDHRG 1 corresponds to results of ideal HRG,
 with same number of particles as used in KM or S-matrix formalism.
 IDHRG (PDG 2016) corresponds to results of ideal HRG model using the hadronic spectrum PDG 
 2016 \cite{Patrignani:2016xqp}.
 Result of ideal HRG model with additional resonances which are yet not confirmed are also
 shown (IDHRG (PDG 2016+)).
 Lattice QCD data of $C_{BS}$ are taken from Refs.~\cite{Borsanyi:2011sw} (WB) 
 and Refs.~\cite{Bazavov:2012jq} (HotQCD).}
\label{fig:CBS}
\end{figure}

The correlation between the strangeness $S$ and baryon number $B$ is a sensitive probe
of the relevant microscopic degrees of freedom. The quantity $C_{BS}$ \cite{Koch:2005vg}
defined as $C_{BS}=-3\chi_{BS}^{11}/\chi_S^2$ is one such observable. For a gas of non-interacting QGP, $C_{BS}=1$ but for a gas of hadrons dominated by kaons and anti-kaons- a light quark is always correlated with its strange 
partner (kaons) or vice versa (anti-kaons) $C_{BS}<1$. However, on the other hand, a system dominated by strange baryons which correlate light quark (anti-quark) with strange quark (anti-quark) and hence have $C_{BS}>1$. Therefore, for 
large baryon chemical potential, $C_{BS}$ could be larger than unity in a hadron gas. Moreover, significant difference between LQCD and ideal HRG has been reported previously \cite{Bazavov:2014xya}. It has been 
argued that such discrepancy can be cured by allowing additional strange hadrons which have not been confirmed but are predicted in various quark models \cite{Alba:2017mqu,Chatterjee:2017yhp}. Fig.~\ref{fig:CBS} shows that the difference between LQCD
and ideal HRG can be accounted by including interaction without invoking any additional hadrons.

\begin{figure}
\begin{center}
\includegraphics[width=0.45\textwidth]{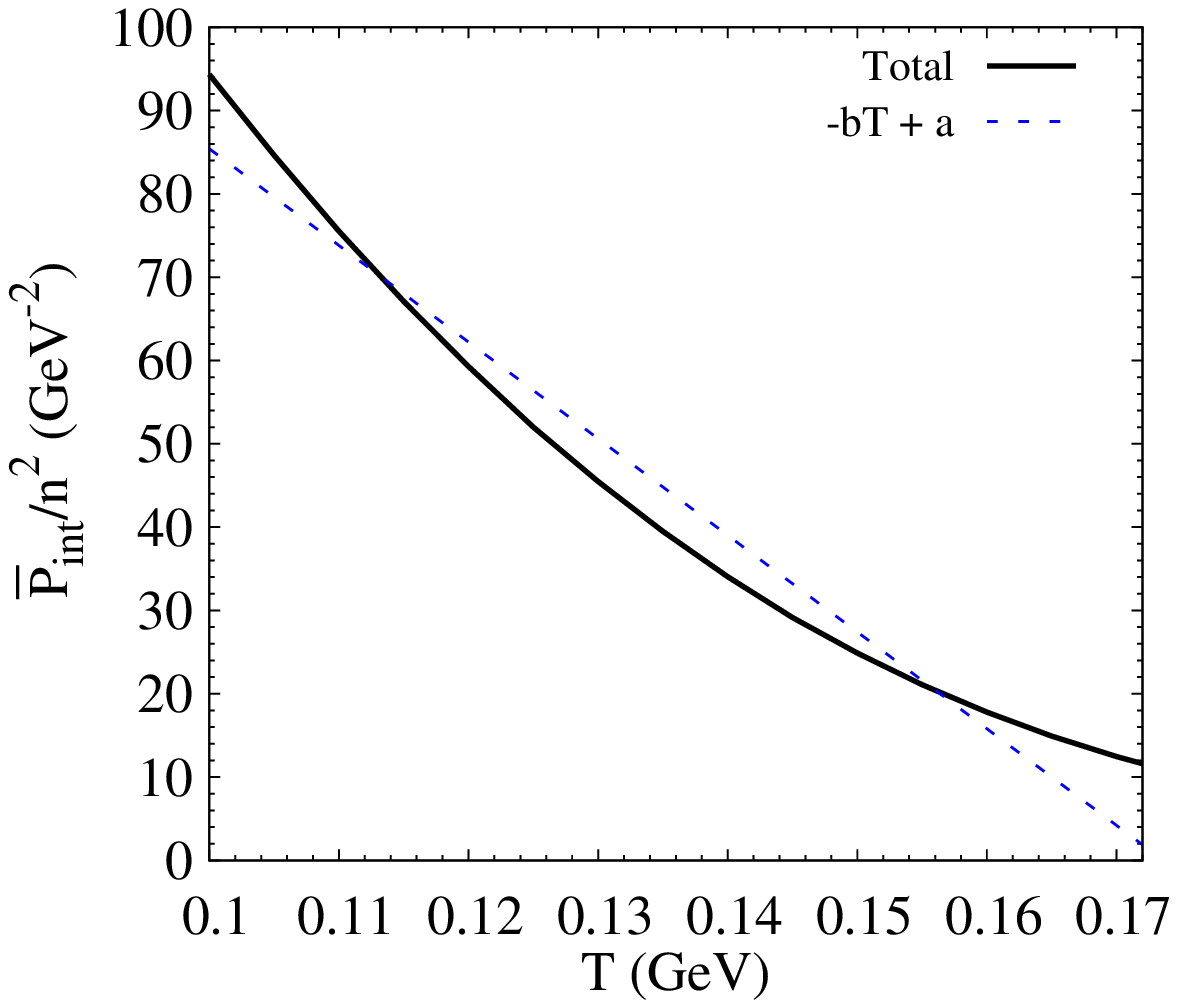}
\end{center}
\vspace{0.5cm}
 \caption{Variation of $\overline{P}_{\text{int}}/n^2$ with temperature at zero chemical potential.
 Total corresponds to both the attractive and repulsive interaction in the HRG model 
 from the current work. The curve is 
 fitted with the
 straight line $-T B(T) = - b T +a$ (see Eq.~\ref{Eq:BT}).}
\label{fig:VDWMatch}
\end{figure}

We match the second virial coefficient obtained using S-matrix formalism with the virial coefficient $B(T)$ of a Van der Waals gas and extract the VDW parameters $a$ and $b$. For a VDW gas 
the coefficient $B(T)$ is given as \cite{Landau:1980mil}
\begin{equation}\label{Eq:BT}
 B(T)=b-\frac{a}{T},
\end{equation}
where $b=16\pi r^3/3$, where $r$ is the hard core radius and $a$ is a positive constant denoting attraction. Thus, the interacting pressure $P_{\text{int}}^{\text{VDW}}$ is related to the 
number density $n^{\text{VDW}}$ for a VDW equation of state as
\begin{equation}\label{Eq:VDW}
 P_{\text{int}}^{\text{VDW}}=-{\left(n^{\text{VDW}}\right)}^2 T B(T),
\end{equation}

Matching $P_{\text{int}}^{\text{VDW}}$ with $\operatorname{\overline{P}_{int}}$, i.e. the second term of Eq.~(\ref{Eq:CanonPressure}) and 
$n^{\text{VDW}}$ with Eq.~(\ref{Eq:Canonnumdensity}), we extract the values of $a$ 
and $b$. Fig.~\ref{fig:VDWMatch} shows the $\operatorname{\overline{P}_{int}}/n^2$ calculated as a function of temperature using S-matrix 
formalism compared to that of an interacting VDW gas. In case of a temperature independent VDW parameters, the curve in Fig.~\ref{fig:VDWMatch} would be a straight line. This study
indicates that the simple (constant) parametrization of the VDW parameters, 
is not correct in a realistic situation, where both the attractive parameter $a$ and 
the repulsive parameter $b$ could in general be temperature dependent. 
This fact also supports models \cite{Venugopalan:1992hy}, where, a temperature dependent
radius was used. 
However, assuming the VDW parameters are temperature independent, a straight line 
fit to the results in Fig.~\ref{fig:VDWMatch} with a functional form of $-bT + a$ 
is carried out to extract the VDW parameters. 
The values of the VDW parameters are 
$a=1.54\pm 0.064$~GeV $\text{fm}^{3}$ and the hard core radius $r=0.81\pm 0.014$~fm. 
We would like to comment here that, the extracted parameters can be seen as some effective values containing
contributions from meson-meson, meson-nucleon and nucleon-nucleon interactions averaged over many hadronic species, 
while Refs.~\cite{Vovchenko:2015vxa, Samanta:2017yhh} extracted these parameters considering only 
baryon-baryon interaction.\par

\section{\label{sec:Summary}Summary}
To summarize, we have included repulsive interaction between hadrons by fitting to 
experimental phase 
shifts which carry the information about the nature of the interaction. The attractive 
part of the interaction are also included and was calculated by parameterizing the 
two body phase shifts using K-matrix formalism \cite{Dash:2018can} which is known to
preserve the unitarity of S-matrix. 
Since the experimental phase shifts for attractive part of the interactions was available 
for the $NN$ scattering, those are used in calculations. 
Thermodynamic quantities like pressure, energy density, trace anomaly, specific 
heat and speed of sound etc. were calculated using the S-matrix formalism. 
The results indicate that the effect of repulsive channels is to suppress the bulk
variables studied here. This finding suggests that contrary to certain channels like $\pi-\pi$ interaction, where the
isospin-weighted sum of $s$-wave attractive and repulsive phase-shifts cancel each other, we found that this
observation is not true for all channels. We find that although some partial cancellation is occurring 
among various phase-shifts in $\pi N$, $KN$ and $NN$ interaction channels, but the resultant interaction is substantial and far from exact cancellation.

Similarly, we compared the Fourier coefficients using S-matrix formalism with lattice data at imaginary chemical potential. 
The leading order coefficient 
$b_1(T)$ reproduces lattice data, while the next to leading order coefficient $b_2(T)$ is smaller than the prediction of 
lattice QCD data. However, we found that $b_2(T)$ is positive for $T<135$ MeV and negative for $T>135$ MeV contrary to 
Ref.~\cite{Vovchenko:2017xad}, which
is negative throughout the temperature range. This can be attributed
to the isospin weighted degeneracy of $NN$ interaction that is positive at lower $\sqrt{s}$ and is negative at higher $\sqrt{s}$.

We found that the most prominent effect of repulsive interactions are seen when we calculate the second and higher order fluctuations and correlation. 
The inclusion of repulsive interaction leads to a better agreement of observables like $\chi^2_B$ and $\chi^2_Q$ with lattice data than the result of only attractive
interaction considered in Ref.~\cite{Dash:2018can} using K-matrix formalism. This is because, in addition to other attractive interactions as considered in ~\cite{Dash:2018can},
resonances like $\Delta (1620)$, $\Delta (1910)$, in the $\pi N$ interaction and resonance like 
$\Sigma(1660)$, in the $KN$ interaction were 
considered attractive in the K-matrix formalism. But here,
we have included such resonances in the repulsive part. This is understood as a comparison to the experimental phase shifts of such resonances
through the factor $\partial\delta^l(\epsilon)/\partial\epsilon$ has rendered that, it is negative and hence repulsive.
Here, we would like to note that the strength from different channels to the repulsive part of the second
virial coefficient is in the order such that $\pi N>KN>NN$. 

Particularly, the two most interesting observations which resulted from the current 
work are as follows. First, we find that
the observable $\chi^2_B-\chi^4_B>0$ for temperatures $T>110$ MeV, in an interacting HRG model discussed in this work, is
contrary to the expectation $\chi^2_B-\chi^4_B=0$ 
for an uncorrelated gas of hadrons like of IDHRG model. 
We also observed that statistics (FD or MB) plays a crucial role on the values and the sign of this observable.
However, the effect of interaction is only 
from $NN$ interaction, which is rather small 
compared to the results obtained by Ref.~\cite{Huovinen:2017ogf,Vovchenko:2016rkn}. The present result can be viewed as a first attempt to 
address such observable in a model which does not have any 
free parameters compared to previous works. This result can be improved by adding other 
baryon-baryon interaction using information from chiral effective theory etc. Second, for the observable 
$C_{BS}$ which is
a sensitive probe of the relevant microscopic degrees of freedom of a system,
the HRG model in the present formalism very well describes the LQCD data.
Also seen from Fig.~\ref{fig:CBS} that IDHRG model with additional 
strange hadrons which 
has not yet been confirmed agrees with the LQCD data at a similar level \cite{Bazavov:2014xya}. 
The difference in physics interpretation is the following: the IDHRG model 
with additional strange hadrons attributes the matching of LQCD data relative to 
normal IDHRG model due to the increase in hadronic degrees of freedom for the system of hadrons. Our 
results in contrast attributes the matching to be due to interactions among the 
constituents that is captured naturally through the formalism used in the current work.

Finally we have tried to quantify the attractive and repulsive interactions in our
model in terms of the VDWHRG attractive and repulsive parameters $a$ and $r$, 
respectively. In doing so we assume that the parameter values do not change with
temperature and the interacting part of the pressure are same in the two models at a given
temperature. It may be noted that our
results as shown in Fig.~\ref{fig:VDWMatch} indicates $a$ and $r$ could be temperature
dependent. We end by saying that as an outlook it would be interesting to calculate various transport co-efficients
in a S-matrix based HRG model and compare to other different types of HRG models and corresponding Lattice QCD results.

\section*{Acknowledgement}
BM acknowledges financial support from J C Bose National Fellowship of DST, Government of India.
SS and AD thank Volodymyr Vovchenko for helpful discussion and 
acknowledge financial support from DAE, Government of India.

\bibliography{RefFile}

\end{document}